# Observation of linear and nonlinear light localization at the edges of moiré lattices

A. A. Arkhipova,[1,2] Y. V. Kartashov,[1,3,*] S. K. Ivanov,[3] S. A. Zhuravitskii,[1,4] N. N. Skryabin,[1,4] I. V. Dyakonov,[4] A. A. Kalinkin,[1,4] S. P. Kulik,[4] V. O. Kompanets,[1] S. V. Chekalin,[1] F. Ye,[5,*] V. V. Konotop,[6] L. Torner,[3,7] and V. N. Zadkov[1,2]

[1]*Institute of Spectroscopy, Russian Academy of Sciences, 108840, Troitsk, Moscow, Russia*
[2]*Faculty of Physics, Higher School of Economics, 105066 Moscow, Russia*
[3]*ICFO-Institut de Ciencies Fotoniques, The Barcelona Institute of Science and Technology, 08860 Castelldefels (Barcelona), Spain*
[4]*Quantum Technology Centre, Faculty of Physics, M. V. Lomonosov Moscow State University, 119991, Moscow, Russia*
[5]*School of Physics and Astronomy, Shanghai Jiao Tong University, Shanghai 200240, China*
[6]*Departamento de Física and Centro de Física Teórica e Computacional, Faculdade de Ciências, Universidade de Lisboa, Campo Grande, Ed. C8, Lisboa 1749-016, Portugal*
[7]*Universitat Politecnica de Catalunya, 08034, Barcelona, Spain*

We observe linear and nonlinear light localization at the edges and in the corners of truncated moiré lattices created by the superposition of periodic mutually-twisted at Pythagorean angles square sublattices. Experimentally exciting corner linear modes in the fs-laser written moiré lattices we find drastic differences in their localization properties in comparison with the bulk excitations. We also address the impact of nonlinearity on the corner and bulk modes and experimentally observe the crossover from linear quasi-localized states to the surface solitons emerging at the higher input powers. Our results constitute the first experimental demonstration of localization phenomena induced by truncation of periodic moiré structures in photonic systems.

PhySH Subject Headings: Solitons; Edge states; Waveguide arrays

Moiré lattices (MLs) are unique physical structures produced by two mutually twisted identical sublattices. For certain twist angles, depending on the symmetry of the underlying sublattices, a ML becomes periodic (*alias* commensurate) while for all other angles it is aperiodic (*alias* incommensurate). Such lattices allow studying physical phenomena enabled by qualitative transformations of their spectra at the crossover from periodicity to aperiodicity. MLs are of paramount importance in the area of low-dimensional materials, such as twisted graphene bilayers, where they led to the discovery of unconventional superconductivity and ferroelectricity [1-3], observation of fractal energy spectrum [4], realization of correlated Chern insulators [5,6], investigation of the edge transport [7,8], and many other effects controlled by twisting of the constituent material layers [9]. MLs have also been studied theoretically in cold atoms, Bose-Einstein condensates, and in hot atomic vapors [10-12].

In optics, the MLs allow elucidation of meta-surfaces with specific reflectivity [13,14] and can be used for realization of the magic-angle lasers [15]. Unlike their material counterparts, photonic MLs can be realized as monolayer structures created by twisted sublattices in a single plane. This enables observation of the localization-delocalization transition for 2D light beams occurring upon variation of the depth or twist angle of the constituent sublattices [16,17], which otherwise had been realized only in 1D geometries [18,19]. The qualitative transformation of linear ML spectrum accompanying the transition from the commensurate to the incommensurate geometries drastically affects self-sustained nonlinear states – lattice solitons. The impact of the geometry in 2D lattice soliton formation has been studied in extended MLs [20,21]. Various photonic emulations of material MLs have been proposed [22,23] aiming to explore flat-band physics and the topological properties of the moiré bands [24] originally discovered in bilayer graphene [25-27].

In this Letter, we present the first example of a photonic fs-laser written ML realized as two mutually twisted at Pythagorean angle square waveguide arrays. The MLs inscription procedure allows its truncation outside a large square region and thus the study of corner and edge excitations in both linear and nonlinear regimes for various orientations of the primary lattice axes with respect to its external edges. We found that despite the bands of such MLs are non-topological, they support in-band linear modes localized predominantly in the deep corner or edge waveguides, with a weak background inside the lattice. Even low-power corner excitations show good localization, in contrast to the fast diffraction observed in the bulk. The existence of such localized corner/edge modes is in sharp contrast with the properties of linear modes in the usual finite 2D arrays composed from identical waveguides [28], where strong localization is not observed, and all modes remain delocalized. Thus, the presence of two sublattices in the ML is crucial for the emergence of the localized states. It also contrasts to the formation of edge and corner solitons, which exist only above a power threshold in truncated 1D [29,30] and 2D arrays [28,31,32] (see [33-35] for reviews), and with the dynamics of edge solitons in photonic graphene [36].

Propagation of the light beam in a medium with focusing cubic nonlinearity and inscribed ML is described by the nonlinear Schrödinger equation for the dimensionless amplitude of the light field $\psi$:

$$i\frac{\partial \psi}{\partial z} = -\frac{1}{2}\Delta\psi - |\psi|^2\psi - \mathcal{P}(\mathbf{r})\psi, \qquad (1)$$

where $\Delta = \partial^2/\partial x^2 + \partial^2/\partial y^2$; $\mathbf{r}=(x,y)$ is normalized to the characteristic scale $r_0 = 10\ \mu\text{m}$; $z$ is the propagation distance normalized to the diffraction length $kr_0^2$; $k=2\pi n/\lambda$; $\lambda = 800$ nm is the working wavelength; function $\mathcal{P}(\mathbf{r}) = p\mathcal{V}[R(\theta_1)\mathbf{r}] + p\mathcal{V}[R(-\theta_2)\mathbf{r}]$ describes the ML created by the superposition of two identical square sublattices $\mathcal{V}[R(\theta_1)\mathbf{r}]$ and $\mathcal{V}[R(-\theta_2)\mathbf{r}]$ with the depth $p = k^2 r_0^2 \delta n/n$, where $\delta n$ is the refractive index contrast (in our case $p \approx 1.95$, that corresponds to $\delta n \sim 2.2\times 10^{-4}$). The sublattices are rotated in the opposite directions, so that the total twist angle between them is equal to $\theta_1 + \theta_2$; $R = R(\theta)$ is the operator of 2D rotation by the angle $\theta$. Each sublattice $\mathcal{V}(\mathbf{r}) = \sum_{l,j} \mathcal{Q}(x - lw_s, y - jw_s)$ represents square array with the period $w_s = 3.3$ (33 $\mu$m) composed of the identical waveguides $\mathcal{Q}(x,y) = e^{-(x/w_x)^2 - (y/w_y)^2}$ that are elliptical due to writing procedure with widths $w_x = 0.25$ (2.5 $\mu$m) and $w_y = 0.75$ (7.5 $\mu$m). Rotation changes orientation of primary axes of the ML, but short and long axes of elliptical waveguides always remain parallel to the $x$ and $y$ axes [see photograph in Fig. 3(a) and a larger version in Fig. S1 in [37]]. When the twist angle between sublattices equals $\theta_1+\theta_2=\arctan[(m^2-n^2)/2mn]$ - the angle associated with the Pythagorean triple $(m^2-n^2, 2mn, m^2+n^2)$ with $m>n>0$, $m,n \in \mathbb{N}$ - the respective infinite ML is exactly periodic or commensurate with the overlapping waveguides at the nodes [see structure in Fig. 1(a),3(a) corresponding to $(m,n)=(2,1)$], while for all other angles it is aperiodic, or incommensurate. Since we are interested in the impact of truncation on the eigenmodes of otherwise periodic MLs, we consider here only the former case. To truncate the ML we keep in both sublattices only the waveguides with coordinates $x_c = w_s(l\cos\theta - j\sin\theta)$, $y_c = w_s(l\sin\theta + j\cos\theta)$, with $l, j$ being integers, $\theta = \theta_1$ and $\theta = -\theta_2$, falling into large square area $(x_c, y_c) \in [-L/2, +L/2]$ of size $L = Nw_s$, where $N$ is an odd integer, whose edges are parallel to $x$ and $y$ axes [see MLs with sizes $L=11w_s$ and $21w_s$ in Fig. 1(a),(b)]. We will vary the angles $\theta_{1,2}$ determining the orientation of the lattice vectors of ML with respect to the external edges of the structure keeping their sum equal to the Pythagorean angle.

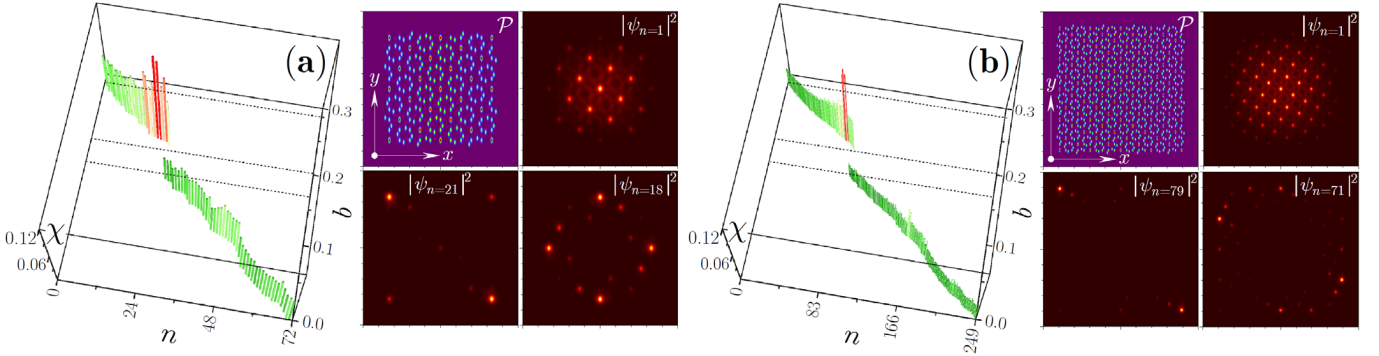

Fig. 1. Form-factors $\chi$ of all linear eigenmodes of truncated ML of size $L=11w_s$ (a) and $L=21w_s$ (b) versus mode index $n$ and propagation constant $b=b_n$. Images to the right of $\chi(n,b)$ dependencies show MLs with $\theta_1 = 0.2048\pi$, $\theta_2 = 0\pi$ and $|\psi_n|^2$ distributions in representative eigenmodes. The lattice and modes are shown within the window $x,y \in [-25, +25]$ in (a) and $x,y \in [-42, +42]$ in (b). Here and below $p=1.95$, $w_s=3.3$.

First we calculate *linear* eigenmodes $\psi = w_n(\mathbf{r})e^{ib_n z}$ of truncated ML, where $b_n$ is the propagation constant of mode with the index $n$ (only discrete spectrum at $b_n \geq 0$ is considered). To characterize the localization we calculate the form-factors $\chi = U^{-1}[\iint |w|^4 d^2\mathbf{r}]^{1/2}$ for all modes (here $w=w_n$), where $U = \iint |w|^2 d^2\mathbf{r}$ is the mode power (higher $\chi$ means better localization). The modes are sorted such that $b$ decreases with increase of $n$, except for the degenerate modes (the explored lattice possesses $C_2$-symmetry). Dependencies $\chi(n,b)$ for ML sizes $L=11w_s$ and $21w_s$ are shown in Fig. 1(a) and 1(b), respectively, for angles $\theta_1 = 0.2048\pi$, $\theta_2 = 0\pi$ when deep waveguides arising due to the overlap of waveguides from different sublattices fall into the corners of the ML. Dashed lines in Fig. 1(a),(b) indicate the edges of the bands for the corresponding infinite ML. Despite all modes of truncated structure fall into the allowed bands, surprisingly, the modes featuring substantially larger form-factors $\chi$ (red bars) appear in the first band that are localized predominantly on deep corner or edge guides with only weak background in the bulk of ML, not even visible on the scale of the figure [see Fig. 1(a),(b)]. With the increase of the ML size [Fig. 1(b)], such modes remain localized, so that the contrast in $\chi$ between them and the delocalized states becomes more pronounced. Such modes are enabled by the ML truncation: their form-factors vary with the angles $\theta_{1,2}$ (defining locations of the deep waveguides at the edges, see Fig. 4). These modes are nontopological (band Chern numbers are 0 at least for 5 highest bands). Considering that the intensity peaks in these modes are well separated in space, to explain their localization we look at the band structure of the underlying infinite lattice. The area of the respective Brillouin zone of the ML, which for the twist angles in Fig. 1 is given by $K^2 = 4\pi^2/5w_s^2$, is notably smaller than that of only one sublattice, $K^2 = 4\pi^2/w_s^2$. Accordingly, the periodic ML is characterized by much flatter bands (see Fig. S2 in [37]). The flatness of the bands results in a significant suppression of the beam diffraction in the ML and in relatively sharp local maxima of its Bloch modes. Since such Bloch modes are predominantly localized around the deepest waveguides, when they are located at the boundaries or corners of the truncated ML, one observes the formation of localized states, as in Fig. 1. Since the band flatness of ML increases with increasing the depth of the sublattices, one also observes a strengthening of the localization of the modes residing in the deep corner or edge waveguides of the truncated structure with increase of $p$.

*Solitons* in truncated MLs can be obtained from Eq. (1) in the form $\psi = w(\mathbf{r})e^{ibz}$ using the Newton method. We found thresholdless soliton families bifurcating from the above in-band localized modes (see e.g. red and green lines in Fig. 2(a) – the soliton power $U$ for these branches vanishes exactly at $b$ values corresponding to the propagation constants of linear modes with indices $n=18$ and $n=21$ from which the bifurcation occurs). These solitons are predominantly localized near the corners or the edges of the ML, but at sufficiently high powers they hybridize with the bulk modes. They are stable at low and moderate powers (at $U \leq 1.36$ for green curve) near bifurcation point. In addition, solitons residing in the corner (magenta curve) and in the center of the structure (blue curve) were found in the semi-infinite gap. The former family corresponds to the surface solitons existing above the threshold power $U \approx 0.59$, while the latter state has no threshold (2D lattice solitons exist above a power threshold in the limiting case of an infinite array in the transverse direction, since in this case they approach delocalized Bloch waves in the cutoff [33], but in our finite ML they bifurcate from $n=1$ mode with finite width $\sim L$). Both these families are stable in the domains with $dU/db > 0$. Since in-band soliton family is stable in the interval of powers exceeding the threshold for surface soliton formation, one may observe a smooth crossover between excitation of the states belonging to these two families for single-site inputs.

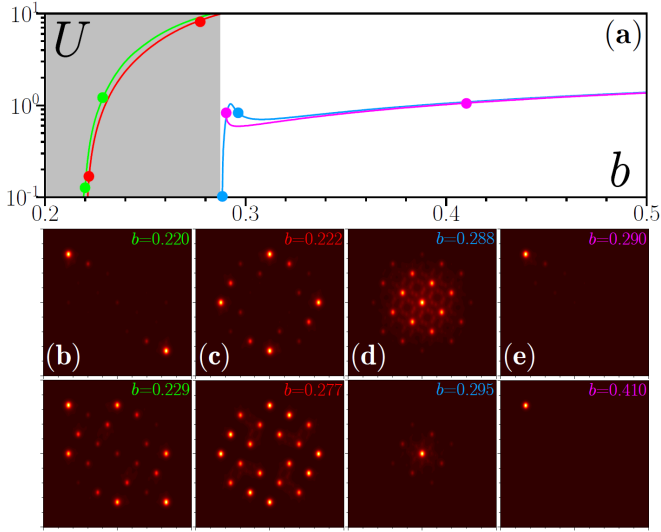

Fig. 2. (a) $U(b)$ dependencies for the soliton families bifurcating from the modes in the allowed band (green and red curves in the gray region) and for the families from the semi-infinite gap bifurcating from the linear limit (blue) and having nonzero excitation threshold (magenta), which are localized respectively in the center and in the corner of the ML with $L=11w_s$, $\theta_1 = 0.2048\pi$, $\theta_2 = 0\pi$. (b)-(e) $|\psi|^2$ distributions in solitons corresponding to the dots in $U(b)$ curves.

In experiment, we inscribed the MLs in 10 cm-long fused silica glass using the fs-laser writing technique (see [37] for details). Microscopic image of one of the fabricated MLs of size $L=11w_s$ is shown in Fig. 3(a). Color circles show typical positions for a single-site excitation (in the corner, at the edge, and in center of ML) for which we used 300 fs pulses of variable energy $E$ from 1 kHz fs Ti:sapphire laser.

Excitation of the waveguide in the center of the ML yields strong diffraction in the linear regime when light reaches lattice edges and even bounces back on the length of our sample corresponding to $z \approx 88$ [see Fig. 3(d) where the blue output experimental patterns for different pulse energies are shown on the left, while theoretical results based on Eq. (1) are shown on the right]. This is because single-site excitation of the central waveguide simultaneously populates many linear modes of the system with comparable weights, most of which are delocalized (see Fig. S3(b) in [37]). With increase of the input pulse energy one observes gradual contraction of the output pattern and eventually formation of well-localized central soliton at $E \sim 540$ nJ [Fig. 3(d)]. In complete contrast, excitation of the deep waveguide in the corner yields much weaker broadening [Fig. 3(e)]. At low energies, the output pattern closely resembles calculated linear corner modes from Fig. 1.

Indeed, such input strongly overlaps with the corner modes leading to the predominant excitation of states localized in the corner, as seen from modal weights in Fig. S3(a) in [37]. All other states, including delocalized ones, are excited with very small weights and do not contribute to the dynamics. Simulations up to the distance of $z = 10^3$ drastically exceeding our sample length also do not demonstrate broadening of the pattern, but one can observe a gradual buildup of light intensity in the opposite corner (as a result of beating between excited corner modes, such as the modes with in-phase and out-of-phase spots in the opposite corners that are nearly equally populated by the single-site input). Even weak nonlinearity arrests this beating leading to the formation of corner state, whose shape remains practically unchanged up to $E \sim 400$ nJ [Fig. 3(e)]. At the higher energies we observe smooth transition to a strongly localized corner soliton [panel with $E = 650$ nJ in Fig. 3(e)]. Localization was not observed for the excitation of shallow waveguides even at powers close to the damage threshold of the material.

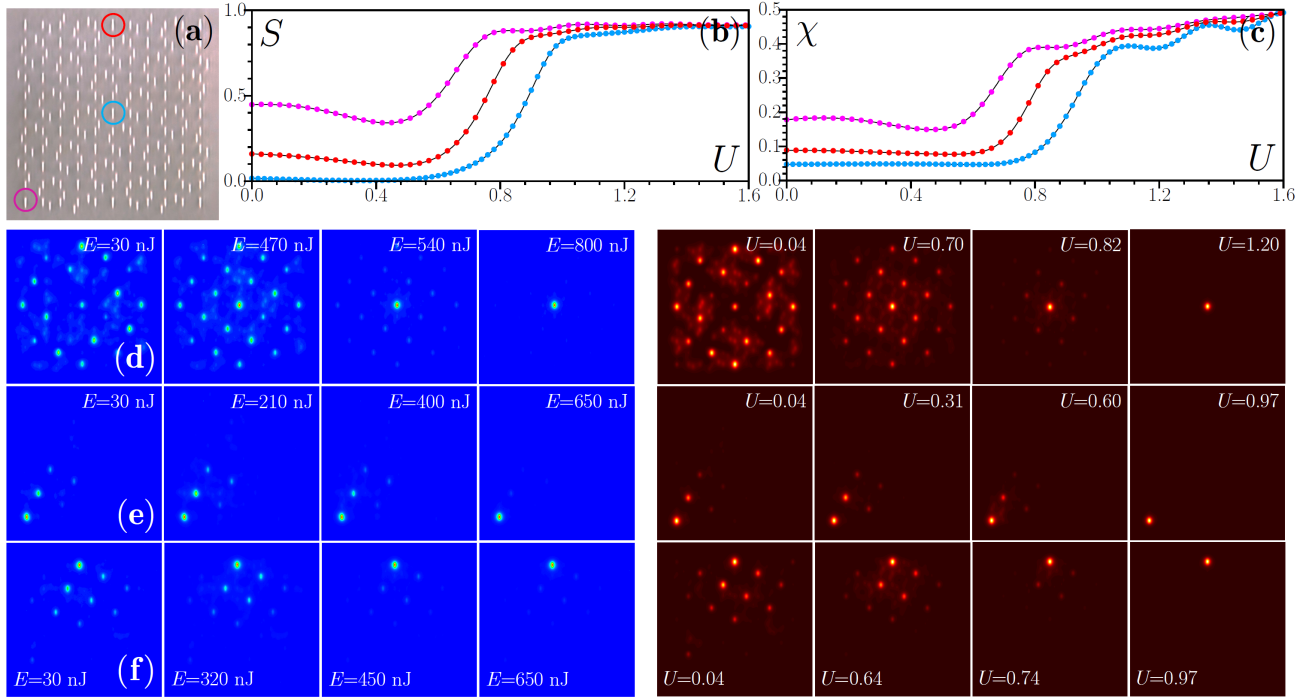

Fig. 3. (a) Photograph of the fs-laser written ML with $\theta_1 = 0.2048\pi$, $\theta_2 = 0\pi$, $L = 11w_s$. Theoretically calculated power fraction, which remains in the excited waveguides (b), and output form-factor (c) vs input power. Color coding for the lines correspond to circles in (a) indicating excited guides. Output experimental (blue images) and corresponding theoretical (dark-red images) distributions for the excitation in the center of ML

(d), in its left bottom corner (e), and in the center of top row (f). Intensity distributions are shown within $x, y \in [-22, +22]$ windows, corresponding to $440$ $\mu$m.

Similar behavior was observed for the excitation of deep waveguide at the upper edge of the ML [see Fig. 3(f)], although the linear output in this case is slightly more extended, in agreement with the shape of the corresponding linear mode [see mode with $n=18$ from Fig. 1(a)]. The power level at which the transition occurs to the strongly localized single-site soliton is larger for the central excitation and lower for the corner one [cf. Figs. 3(d),(e)]. This is confirmed by theoretically calculated dependence [Fig. 2(b)] of the power fraction $S = U_{\text{out}}/U_{\text{in}}$ that remains at the output within the circular area of radius $w_s$ surrounding excited waveguide on the input power $U$. The transition to large $S$ values indicating on strong localization occurs at notably different powers for different excitations. For the corner excitation $S$ takes sufficiently large values even at $U \to 0$, signalizing on the efficient excitation of the linear corner modes. Similar conclusions can be drawn from the dependence of the output form-factor $\chi = U^{-1}[\iint |w|^4 d^2\mathbf{r}]^{1/2}$ of the field on $U$ from Fig. 3(c).

To study how localization properties in the truncated MLs depend on the relative rotation of the ML with respect to the truncating frame, we inscribed several lattices of the same size $L = 11w_s$ for different sets of angles corresponding to the fixed $\theta_1 + \theta_2 \approx 0.2048\pi$ and analyzed their corner and edge excitations. Left column in Fig. 4 shows photographs of different lattices and excited waveguides.

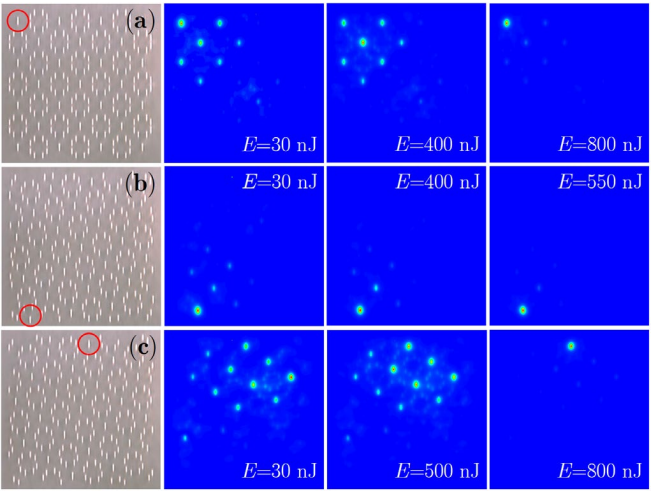

Fig. 4. Photographs of the fs-laser written MLs with different orientation with circles indicating excited waveguide (first column) and measured output intensity distributions for different input pulse energies (second-fourth column) for (a) $\theta_1 = 0.1024\pi$, $\theta_2 = 0.1024\pi$, (b) $\theta_1 = 0.1536\pi$, $\theta_2 = 0.0512\pi$, (c) $\theta_1 = 0.1639\pi$, $\theta_2 = 0.0409\pi$.

Localization properties strongly depend on $\theta_{1,2}$. Thus, in configuration with $\theta_1 = 0.1024\pi$, $\theta_2 = 0.1024\pi$, which is highly symmetric because the primary axes of the ML are oriented exactly along the diagonals of truncating square region, corner mode shows weaker localization in the linear case [Fig. 4(a)]. At intermediate energy levels $E \sim 400$ nJ one can even observe nonlinearity-induced delocalization, when intensity maximum shifts into ML bulk. In contrast, for $\theta_1 = 0.1536\pi$, $\theta_2 = 0.0512\pi$ we have ML with only one overlapping waveguide located in close proximity of each edge, for which we observed even better corner localization than in Fig. 3, practically in single waveguide [Fig. 4(b)] that becomes more pronounced with increase of $E$. Finally, when $\theta_{1,2}$ are such that all overlapping waveguides are shifted slightly into the volume of the structure, as it occurs for $\theta_1 = 0.1639\pi$, $\theta_2 = 0.0409\pi$, one observes strong diffraction for the edge excitations signalizing on the absence of modes localized near the edges [Fig. 4(c)]. Considerable nonlinearity is then needed to achieve the edge localization.

In conclusion, we demonstrated that the truncation of otherwise periodic photonic MLs introduces fundamentally new physical phenomena, and in particular may lead to the appearance of in-band modes localized predominantly near the corner or edge waveguides. Our observations pave the way to the exploration of unconventional localization phenomena in various physical settings such as low-dimensional materials, twisted bilayer graphene, atomic systems, and acoustics. In particular, the crossover to strongly localized corner, edge and bulk solitons at high powers may occur in other moiré heterostructures driven in the nonlinear regime.

*kartashov@isan.troitsk.ru
*fangweiye@sjtu.edu.cn